\documentclass[10pt,amssymb,twocolumn,notitlepage]{article}

\usepackage{epigraph}
\usepackage[affil-it]{authblk}
\usepackage{dcolumn}
\usepackage[stable]{footmisc}
\usepackage{blkarray}
\usepackage{bm}
\usepackage[font=scriptsize]{caption}
\usepackage{subcaption,graphicx}
\usepackage{amsmath}
\usepackage[table]{xcolor}
\usepackage[margin=1.7cm]{geometry}
\usepackage{tcolorbox}
\usepackage{hyperref}
\usepackage{gensymb}
\usepackage[font=footnotesize]{caption}
\usepackage{hyperref}
\usepackage{amssymb}

\title{Modeling Consonance and Its Relationships with \\ Temperament, Harmony, and Electronic Amplification }
\author{Luciano da Fontoura Costa \\ \emph{luciano@ifsc.usp.br}}
\affil{S\~ao Carlos Institute of Physics -- DFCM/USP} 

\begin{document}

\twocolumn[
\begin{@twocolumnfalse}
    \maketitle
    \begin{abstract}
    After briefly revising the concepts of consonance/dissonance, a respective
    mathematic-computational model is described, based on Helmholtz's 
    consonance theory and also considering the partials intensity.  It is then
    applied to characterize five scale temperaments, as well as some minor and major
    triads and electronic amplification.  
    In spite of the simplicity of the described model, a surprising agreement is often
    observed between the obtained consonances/dissonances and the typically
    observed properties of scales and chords.  The representation of temperaments
    as graphs where links correspond to consonance (or dissonance) is 
    presented and used to compare distinct temperaments, allowing the identification
    of two main groups of scales.  The interesting issue of nonlinearities in electronic
    music amplification is also addressed while considering quadratic distortions, and
    it is shown that such nonlinearities can have drastic effect in changing the 
    original patterns of consonance and dissonance. 
    \end{abstract}
\end{@twocolumnfalse} \bigskip
]

\setlength{\epigraphwidth}{.49\textwidth}
\epigraph{`Geometry is in the humming of strings.'}{Pythagoras.}

\section{Introduction}

Sound and music are intrinsically human concepts and activities.  In other
words, there are no definitions to be found in nature regarding what sound and
music are, or how they should be.  These definitions and properties are
\emph{relative} to us, humans.  As such, these concepts are inherently subjective,
and they may vary in time and space.  Indeed, the history
of music has continuously witnessed big changes in instruments, styles, and theories,
contributing to ever increasing diversity of results.

Nevertheless, there is an aspect somewhat stable in sound and music,
and this has to do with \emph{consonance} and \emph{dissonance}.  
These concepts have permeated the history of tonal music, while
also playing important roles in many other musical traditions.  
Basically, consonance has to do with our sensation
of `harmony' between two simultaneous sounds, or notes.  As such, 
consonance/dissonance is at the very basis of \emph{sound combinations} 
along time and space, which underlies \emph{music} . 

One of the first mathematical approaches to consonance was developed by
Pythagoras (e.g.~\cite{Kahn:2001}), by considering ratios between the length of strings.
Afterwards, the historic records mostly go back to monophony, especially
Gregorian chant.  With time, drones were added, initiating a very simple
kind of harmony.  Subsequent developments would lead to two parallel 
voices, contrary/oblique voices, and then to ever increasing types of polyphony.
With the incorporation of more and more voices, the problem of combining 
sounds in time and space became ever more critical.

Consonance has played an important role in many musical traditions.  In Indian
classical music (e.g. Bharata), for instance, discussions on consonance seem
to go back at least to the $VIII$ century $bC$, involving the concepts of \emph{vadi} (`sonant'), 
\emph{samvadi} (`consonant'), \emph{vivadi} (`dissonant') and \emph{anuvadi} (`assonant')
(e.g.~\cite{indian_music})

To make a long (and interesting) history short, the progression to polyphony followed,
or even implied, the development of \emph{harmony} and \emph{counterpoint},
which can be informally understood as the activity of combining simultaneous
or subsequent sounds, respectively (e.g.~\cite{Laitz:2009, Stone:2018}). 
One of the first more formal approaches
to consonance was developed by J. P. Rameau, who argued strongly
against the current view overlooking differences between consonance and dissonance.
To him, consonance was a kind of \emph{repose}, so that dissonance would imply
a depart from that state (e.g.~\cite{Christensen:1993}).

Because of its central role in sound and music, consonance motivated
several theories, such as based on interval ratios or interaction between
harmonics/partials.  In particular, the latter was developed mainly
by H. L. F. Helmholtz (1821--1894).  Backed by careful experiments,
Helmholtz advanced the idea that consonance derives from interrelationships
between partials, producing constructive combination of frequencies or, alternatively, 
potentially unpleasant \emph{beats} (e.g.~\cite{Cunningham}).

Recall that, given a fundamental tone, the accompanying higher frequency
tones are called \emph{partials}.  In case these partials follow the harmonic
series (i.e. string wavelengths of~$1, 1/2, 1/3, 1/4,\ldots$), they are called 
\emph{harmonics}.  

In this text (a previous version of which appeared recently~\cite{CostaCons}), 
we approach the interesting issues of consonance/dissonance
by considering Helmholtz's respective theory.  More specifically, we derive
a simple mathematic-computational model (e.g.~\cite{CostaModeling}) 
capable of, given two sounds, providing
degrees of consonance and dissonance.   In addition to considering the
frequencies of the involved partials, the respective amplitudes are also taken
into account their magnitudes.  Then, we use this model to revisit some of the 
main temperaments
(informally speaking, ways of assigning frequencies to notes)
as well as some aspects of basic harmony, especially triad consonance.

The interesting and important issue of nonlinearity in electronic amplification 
is also addressed with respect to quadratic effects, and the results show that
nonlinear amplification can substantially change the original patterns of
consonance and dissonance found in the original sounds.

\section{On the Mixture of Sinusoidals with a Same Frequency}

A first interesting issue to be addressed regards what happens when two sinusoidal
sounds with a \emph{same} frequency $\omega = 2 \pi f$, but distinct amplitudes 
($a_1$, $a_2$) and
phases ($\theta_1$, $\theta_2$), are combined, i.e. $a_1 cos(\omega t + \theta_1) + 
a_2 cos(\omega t + \theta_2)$.  

\begin{figure}[h!]
\centering{
\includegraphics[width=4cm]{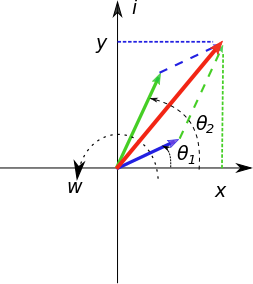}  
\caption{Phasor representation of two sinusoidal sounds with the same
frequency $f = 2 \pi f$ but distinct amplitudes ($a_1$, $a_2$) and
phases ($\theta_1$, $\theta_2$).  Because these signals have the same
frequency, this can be disregarded, so that the phasors consider only the
respective phases, yielding $a_1 e^{i \theta_1}$ (shown in blue)
and $a_2 e^{i \theta_2}$ (shown in green).
The addition of these two sounds can be implemented vectorially in the
complex plane (and then taking the real projection), yielding the phasor shown in red. }
\label{fig:phasors}}
\end{figure}

This interesting question can be simply addressed by considering the phasor representation
(e.g.~\cite{CostaPhasor}) of these two sounds, respectively $a_1 e^{i \theta_1}$ and
$a_2 e^{i \theta_2}$, as illustrated in Figure~\ref{fig:phasors}.

Phasor representation allows us to add, in complex or vector fashion, the
involved phasors, resulting the phasor shown in red in Figure~\ref{fig:phasors},
which has real and imaginary parts, as well as its magnitude $a$ and
phase $\theta$, easily calculated as

\begin{eqnarray}
  x = a_1 cos(\theta_1) + a_2 cos(\theta_2) \nonumber \\
  y = a_1 sin(\theta_1) + a_2 sin(\theta_2) \nonumber \\  
  a = \sqrt{x^2 + y^2}  \nonumber  \\
  \theta = atan\left( \frac{y}{x} \right).  \nonumber
\end{eqnarray}.

So, we have that
 
\begin{equation} 
   a_1 cos(\omega t + \theta_1) + a_2 cos(\omega t + \theta_2) =
    a \, cos\left(\omega t  \theta \right).   \label{eq:two_cos} 
\end{equation}

Mixtures of sines and cosines can also be handled by using 
Equation~\ref{eq:two_cos} while recalling that  $sin(\alpha) = 
cos(\alpha - \pi/2)$

Thus, the linear combination of any number of sines and cosines with the
same frequency, no mattering how different are their magnitudes and phases are,
will always yield a simple cosine with effective magnitude $a$ and
phase $\theta$ as a result.  In the next section we show that more
interesting results arise when linearly combining two or more sinusoidals
with \emph{different} frequencies.

\section{Mixture of Sinusoidal Signals with Distinct Frequencies}

Given two sinusoidal tones with respective distinct frequencies $\omega_1 = 2 \pi f_1$ 
and $\omega_2 = 2 \pi f_2$, $\omega_1 < \omega_2$, it immediately 
becomes interesting to quantify the
relationship between these frequencies.  One possibility is in terms of their 
\emph{ratio} $f_2/f_1$, and another is in terms of the measurement called
\emph{cent}, which is defined as

\begin{equation}
   cent(\omega_2, \omega_1) = cent(f_2,f_1) = 1200 \, log_2\left(\frac{w_2}{w_1} \right).
\end{equation}

Observe that 100 cents equate a equally-tempered \emph{half-tone}, corresponding
to each subsequent note along a scale.

Let's sum these two signals as

\begin{equation}
   s(t) = a_1 cos(\omega_1 t + \theta_1) + a_2 cos(\omega_2 t + \theta_2).   \nonumber
\end{equation}

\begin{figure}[h!]
\centering{
\includegraphics[width=8cm]{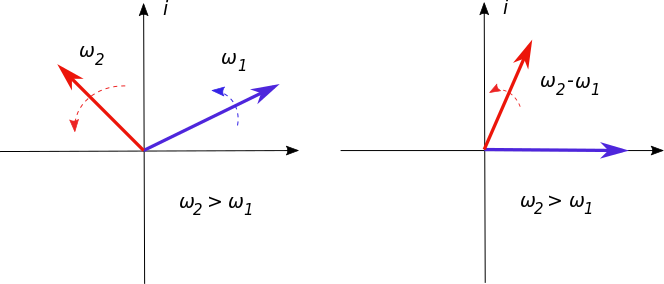}  
\caption{(a) Phasor representation of two sinusoidal functions with distinct
frequencies, amplitudes and phases. (b) It is possible to keep the phasor
corresponding to the smaller frequency signal parallel to the real axis while
analyising the behavior of the other phasor, which now rotates with angular
velocity (or frequency) $\omega_2 - \omega_1$.}
\label{fig:two_ws}}
\end{figure}

Again, the phasor representation turns out to be particularly useful in analysing
this signal addition.  Figure~\ref{fig:two_ws} illustrates this situation.

Because the two frequencies are now different, for simplicity's sake we can
keep the slower-rotating phasor (shown in blue) parallel to the
real axis.  Observe that the original phases $\theta_1$ and $\theta_2$ are
therefore lost, but this does not matter for our following analysis.

\begin{figure}[h!]
\centering{
\includegraphics[width=8.8cm]{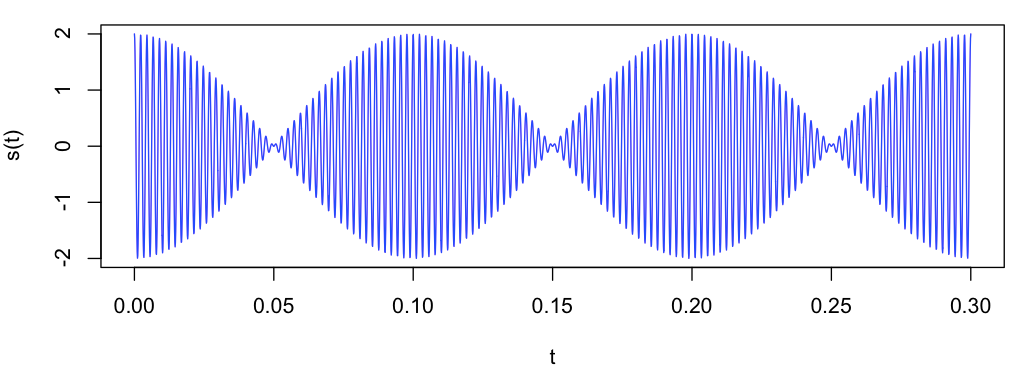}  
\caption{Signal resulting from the sum of two sinusoidals $f_1=440Hz$ and $f_2=450Hz$.
   Observe the amplitude \emph{modulation} taking place with frequency $10Hz$.  Both
   original signals have unit amplitude.}
\label{fig:beats}}
\end{figure}

Observe that the second phasor (shown in red) now rotates with 
angular velocity  $\delta \omega = \omega_2 - \omega_1$.  So, as it moves, the magnitude
of the resulting phasor (projection onto the real axis) 
varies from a maximum value of $a_1$, for $\delta \omega =0$,  
to its smallest value, observed at delta $\delta \omega  =\pi$ .   This oscillating 
effect on the amplitude takes place with 
angular frequency $\omega_2-\omega_1$.  Figure~\ref{fig:beats} illustrates
the amplitude oscillation produced by the addition of two signals with 
respective frequencies $f_1=440Hz$ and $f_2 = 450Hz$ and identical 
unit magnitudes, so that a beat frequency of $10Hz$ is obtained.

It is interesting to observe that, though there is a difference of angular velocities,
the sum of the two signals only incorporates the two original frequencies
$\omega_1$ and $\omega_2$, so that $\omega_2 - \omega_1$ does not
belong to the respective frequency spectrum.  This can be readily appreciated 
by taking into account this addition in the Fourier domain, where only these two 
frequencies can be present.  

Now, we come to an interesting issue: the way in which \emph{humans} perceive this
frequency difference \emph{also depends on its magnitude}.  For $\delta \omega$ 
very small, (e.g.~$<2Hz$, approximately), we tend to perceive the two sounds as a single,
\emph{tuned} note.  For slightly larger difference values (typically $10Hz > \delta \omega
\geq 2Hz$),  a single frequency with a relatively slow varying amplitude tend to be heard.  
Indeed, difference in this range are smaller than our minimal frequency limit of $20Hz$.  

For larger values of $\delta \omega$ (e.g.~$60Hz > \delta \omega > 10 $), the
added sinusoidal signals tend to cause an unpleasant sensation, as the resulting
sound cannot be distinguished between a single or double frequency
stimulus.  The  two latter situations are often described as the \emph{beat}
effect, and are typically avoided in music and instrument construction and tuning.  
Recall that this phenomenon, jointly with tuned sounds, provided some of the
main motivation for Helmholtz's consonance theory.

The situation where $\delta _\omega > 60Hz$ typically leads to two separate sounds,
which are typically called \emph{intervals}.  We will discuss this concept further in the
next section.

\section{Intervals}

Table~\ref{tab:ints} presents the main intervals often considered in tonal 
music, as well as the respective just ratio (in cents) and equal-temperament
counterpart.  The type of consonance typically perceived for each case is
also provided in the last column of this table.

\begin{table}
\caption{The intervals characteristic of tonal music, with respective ratios
and types of consonance/dissonance. } \label{tab:ints} 
\begin{center}
\begin{tabular}{|  c  || c  | c |  c |}
\hline
 interval  &  just ratio  &  just cents  &  type of cons.  \\
 \hline
unison              &  1/1   & 0  & absolute cons.  \\
perfect octave  &  1/2   & 111.73  &  absolute cons.  \\
fifth                   &   2/3  &  203.91 &  perfect cons.  \\
forth                  &  3/4  &  315.64 &   medium cons.  \\
major sixth       &  3/5   &   386.31  &   medium cons.  \\
major third       &  4/5   &   498.04  &   medium cons.  \\
minor third       &   5/6  &  582.51   &   imperfect cons. \\
minor sixth       &   5/8  &  701.96  &   imperfect cons. \\
major second  &    8/9  &   813.69   &   diss. \\
major seventh   &  8/15  &  884.36    &   diss.\\
minor seventh   &  9/16  &   996.09   &   diss.\\
minor second   &  15/16  & 1088.27     &   diss.\\
tritone               &  32/45  &   1200   &   diss.\\
\hline
\end{tabular}
\end{center}
\end{table}

It should be observed that this classification of consonances is not
definitive, and some of these intervals have been perceived 
differently along time and space.   This may have to do with the fact
that consonance, as we will see further on this text, depends on
the partial features of each type of sound, as well as on the type
of assumed temperament.  

The just intonation, commonly used to tune some types of instruments,
represents the basis for some temperament systems, such as the
pythagorean.

Going back to the previous table, we have that only three or four
intervals -- namely those corresponding to unison, octave, and 
fifth -- are understood as presenting more definite consonance.  It is thus
hardly surprising that the fifth, and to a lesser extent the forth,
represent such an important reference in occidental music composition.

 Four intervals are understood as having intermediate consonance,
 and the remainder five are typically understood to be dissonant, with
 emphasis on the tritone (e.g. $C-F\#$).
 
 Classification of consonance/dissonance has traditionally relied on
 relatively subjective human appreciation.  In the following we will describe
 Helmholtz's experimental/mathematical approach to consonance and
 then consider it for developing a simple, and yet relatively effective,
 model capable of automatically estimating degrees of consonance and
 dissonance.

\section{Helmholtz's Theory of Consonance, and a Simple Model}

Helmholtz's theory of consonance was presented mainly in the second part
of his \emph{Sensations of Tone} (e.g.~\cite{Cunningham}).   The two involved sounds
are understood in terms of their respective harmonic series, and these
are pairwise compared in order to identify respective tuning and beats.
The resulting sensation of consonance/dissonance would be related to
the intensity in which these constructive interactions and beats occur
given each the two original sounds.  Interestingly, we have that two pure
sounds (i.e. without partials) could hardly be compared regarding their
consonance.

Here, we consider Helmholtz's main concepts to develop a simple 
mathematic-computational model, capable of estimating in a quantitative way 
the consonance/dissonance between two sounds $s_1(t)$ and $s_2(t)$
with respective fundamental frequencies $f_{0,1}$ and $f_{0,2}$.

A first important issue regards the frequency distribution of partials.  For
simplicity's sake, and without loss of generality, we assume that these
follow the \emph{harmonic series}, leading to partials of $f_0, 2f_0, 3f_0, 4f_0, 
\ldots$.  This is not to say that real sounds necessarily follow this structure,
but other partial models can be immediately considered in the described model.

\begin{figure}[h!]
\centering{
\includegraphics[width=8.8cm]{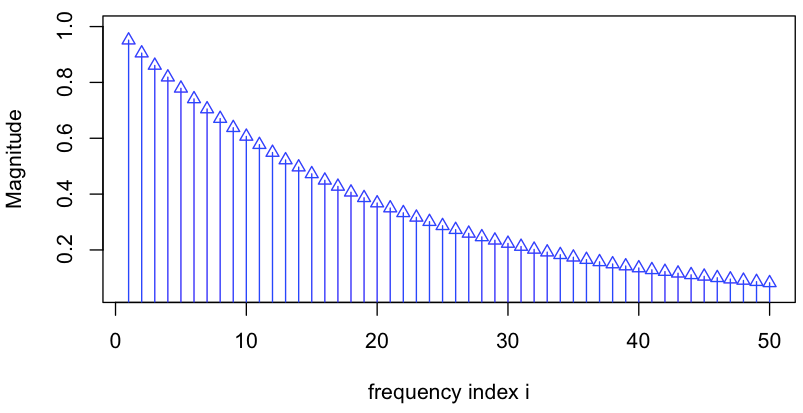}  
\caption{The adopted negative exponential way in which the magnitude of the
    harmonic partials are assumed to decay with the frequency index $i$
    (respective to fundamental $f_0 = 1Hz$).}
\label{fig:decay}}
\end{figure}

As there are virtually infinite harmonics, we also have to delimit the frequency
extent of these sounds.  For this purpose, we consider
the fact that the partials of most sounds decay steadily with frequency
(and also along time), so that too high
harmonic partials result inaudible.  For simplicity's sake, we consider that the partials
decay in terms of a negative exponential.   Figure~\ref{fig:decay} illustrates
the adopted partial decay with respect to the frequency index $i$ corresponding
to $f_0 = 1Hz$.

Though other configurations can be adopted, we henceforth assume a partial
decay scheme as in Figure~\ref{fig:decay}, incorporating a maximum of $N= 50$
harmonic partials (observe that the limits of human frequency perception, varying
roughly as $20Hz \leq f \leq 20kHz$, should also be considered).

Now, given two signals represented in terms of their partial content, we need to
devise some means to compare these partials in pairwise fashion and then
estimate, for each case, the respective degrees of consonance and dissonance.

\begin{figure}[h!]
\centering{
\includegraphics[width=8cm]{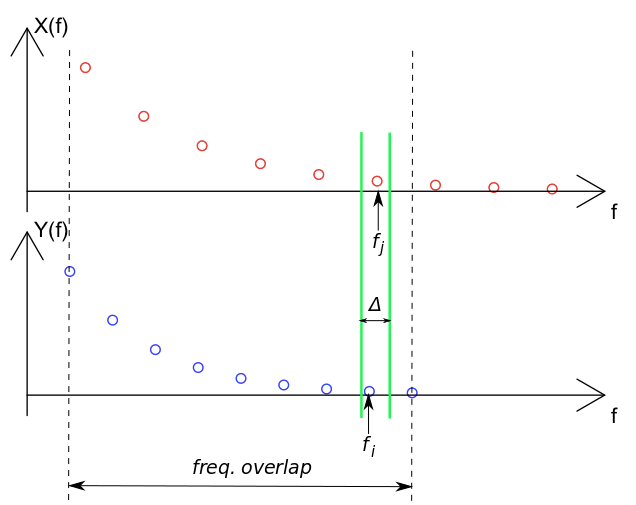}  
\caption{The spectra of the two sounds to be compared regarding the consonance/dissoance
of the respective partials.  The lowest frequency fundamental sound is taken as a reference
for the comparison, which involves both the frequency difference as well as the
respective magnitudes.}
\label{fig:two_specs}}
\end{figure}

Figure~\ref{fig:two_specs} illustrate the spectrum of two sounds to be compared $X(f)$ and
$Y(f)$ (red and blue, respectively).  Observe that $X(f)$ has higher 
fundamental frequency, implying in more widely spaced partial intervals as well
as effective frequency extent.  

Let's assume the lower fundamental signal $Y(f)$ to be our reference.  For each
of its partials, with respective frequency $f_i$, the closest partial $f_j$  belonging to the
other signa, $X(f)$, is identified.  The frequency difference $\delta f_{i,j} = | f_j - f_i |$ is
then taken, being then classified as: (i) consonance if $\delta f_{i,j} < f_c $; 
(ii) dissonance if $f_c \leq \delta f_{i,j} < f_d $; and (iii) neutral for $\delta f_{i,j} \geq f_d$.

So, many consonance or dissonance matches, as well as several neutral cases,
can be identified for a same pair of sounds.  In order to derive an overall degree
of consonance and dissonance between the two original tones $X$ and $Y$, we can 
respectively apply

\begin{eqnarray}
    consonance(X,Y) = \sum_{cons. (i,j)} M(i) \, M(j) \, \delta f_{i,j}  \label{eq:cons} \\
    dissonance(X,Y) = \sum_{diss. (i,j)} M(i) \, M(j) \, \delta f_{i,j}  \label{eq:diss}
\end{eqnarray}

Observe that the sum takes place over the pairs $(i,j)$ of consonant harmonic
partials in the case of Equation~\ref{eq:cons}, and dissonant pairs of partials
in Equation~\ref{eq:diss}.  Other ways of combining the partial relationships
are possible and can be immediately adapted into the developed model.

Number theory, as well as the convolution mathematical operation,
can also be applied to identify the potential consonances and
dissonances arising from distinct partial distribution and specific sound intervals.

\begin{figure}[h!]  \
\centering{
\includegraphics[width=8cm]{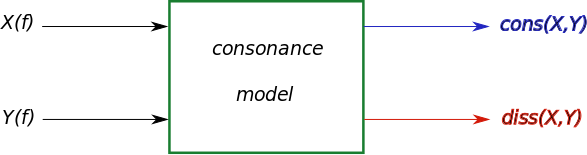}  
\caption{A `\emph{consonance machine},' incorporating a simple model
based on Helmholtz's respective theory. The two sounds to be combined
are input into the system, which provides degrees of consonance and
dissonance as output.}    \label{fig:model}  }
\end{figure}

So, we end up with a model that, as illustrated in Figure~\ref{fig:model}, is
capable of automatically assigning degrees of consonance or dissonance
to pairs of presented sounds represented in terms of partials and magnitudes
(spectral).

Observe that, according to the adopted assumptions, 
the same pair of sounds (interval) has intrinsic levels of both consonance
\emph{and} dissonance (this happens rarely, though).  It is also possible to adopt a threshold for classifying
an interval as being consonant/dissonant.  In the following we adopt 5 and 4
as thresholds for consonance and dissonance.  An exponential decay of $0.08$,
as well as $f_c = 10Hz$ and $F_d = 60Hz$ have been also considered, as these
tended to yield more appropriate results.

Some interesting implications can be observed about the above outlined approach to
consonance.   First, we have that consonance tends to decrease for larger
difference of fundamental frequencies because of resulting smaller overlap
between the partials.     Then, we also have that even linear filtering can influence
the resulting consonance/dissonance, as a consequence of changes in the partial
amplitudes.  So, for instance, low-pass filtering can potentially contribute to 
reduction of dissonance as a consequence of the respectively implied attenunation
of higher partials.

All in all, we also have that consonance/dissonance ultimately depends on a series
of factors, including temperament type, distribution of partials (other than harmonic
series), and different decay profiles of partial magnitudes.  

Nevertheless, the developed model
allows us, to some limited degree of precision that also depends on the configuration
of the involved parameters and roles (which can be adapted), to study several
issues in sound and music.  In the remainder of this text, we apply the simple
developed model to investigate temperaments and basic harmony, especially
triads.

\section{Temperaments}

Because scales are so important in music, several such systems have been
proposed along centuries.  Here, we will be limited to five representative
cases: (i) equal temperament; (ii) pythagorian; (iii) just major; (iv) mean-tone;
and (v) Weckmeister.  

\begin{figure}[h!]
\centering{
Equal Temperament:  \\
\includegraphics[width=8cm]{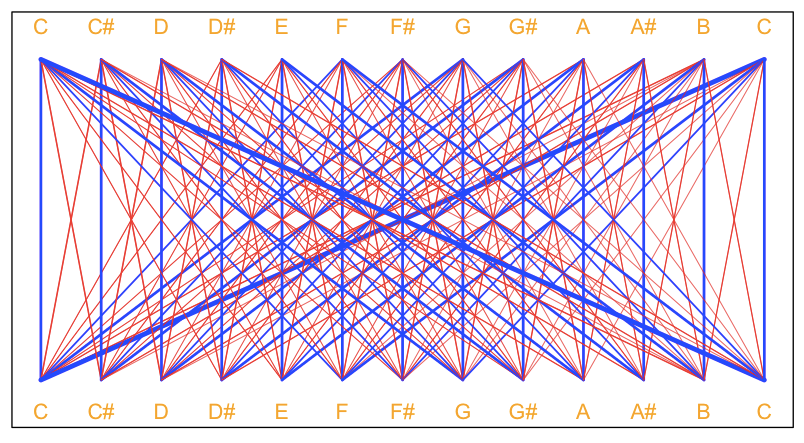}  
\caption{The consonance/dissonances obtained for the equal temperament
scale starting at central $C$ by using the described model.  The consonances
are shown in blue, while the dissonances are presented in red.  Intervals
with higher levels of consonance and dissonance are shown in brighter colors. The 
consonance/dissonance intensities are reflected in the respective edge widths.  
Observe the almost
perfectly uniform pattern of consonances/dissonances obtained for the
equal temperament approach.}
\label{fig:sc_equal}}
\end{figure}

Figure~\ref{fig:sc_equal} illustrates the bipartite graph of consonance (blue)
and dissonance (red) relationships between the 12 tones.   This scale is
characterized by having fixed relationship between each subsequent
half-tone, in the sense that $f_{i+1}/f_i = 2^{1/12}$.  Therefore, any
minor or major modes will be characterized by identical intervals,
providing uniform subsidy for modulations and transpositions.

The effect of the uniform interval characteristic of this type of scale becomes
evident in this figure.  Except for the fact that the consonance tends to change
as one of the two frequencies becomes larger than the other,
the obtained consonance/dissonance patterns are mostly similar for every
note, confirming the properties often expected from this scale.  However, observe
that the interval relationships are mostly dissonant, with the main exceptions
of the major forth and major fifth.  So, interval uniformity is achieved at the
expense of consonance.

\begin{figure*}[h]  
\centering{
        (a) Pythagorean:   \hspace{6.5cm}  (b) Mean-tone:  \\
        \includegraphics[width=8cm]{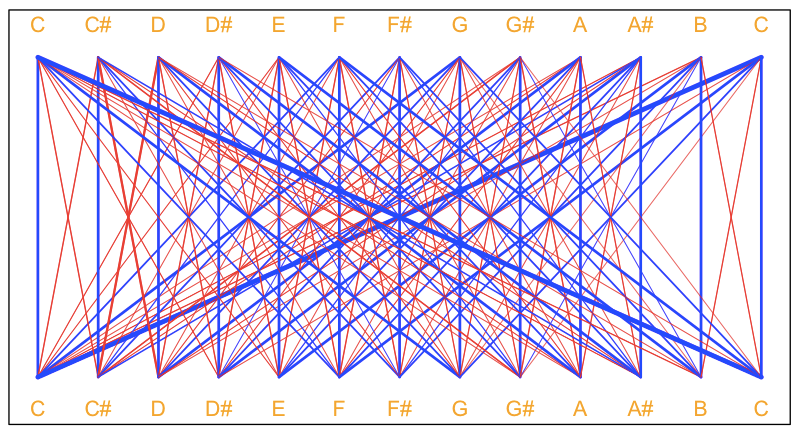}   \hspace{1cm} 
        \includegraphics[width=8cm]{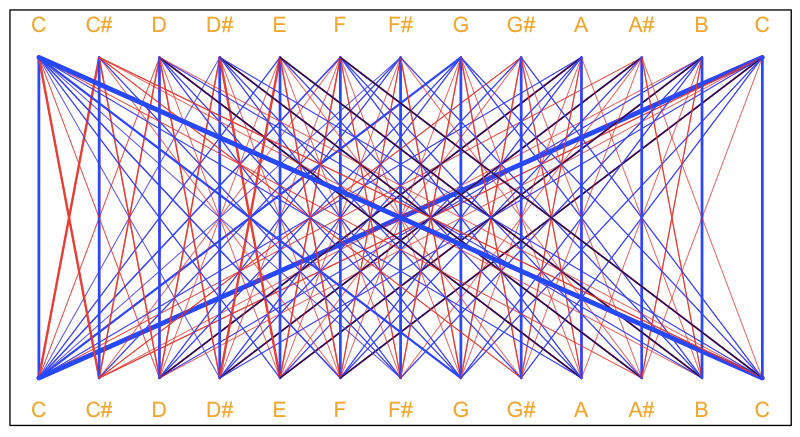}  \\  \vspace{0.5cm}
        (c) Just Major: \hspace{6.5cm}  (d) Werckmeister: 
        \includegraphics[width=8cm]{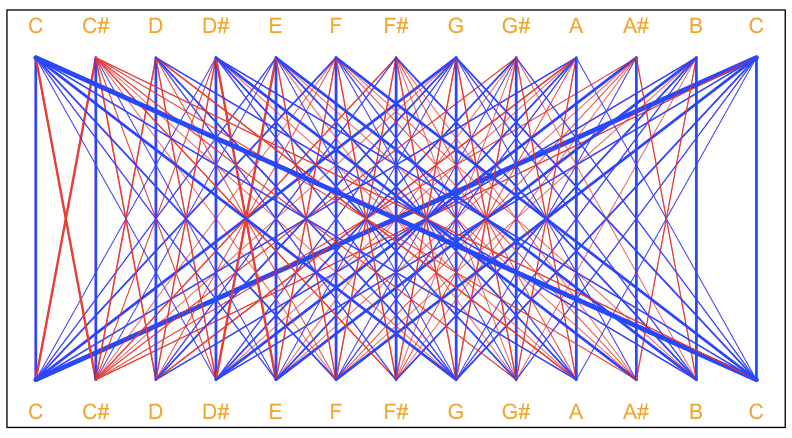}  \hspace{1cm}
        \includegraphics[width=8cm]{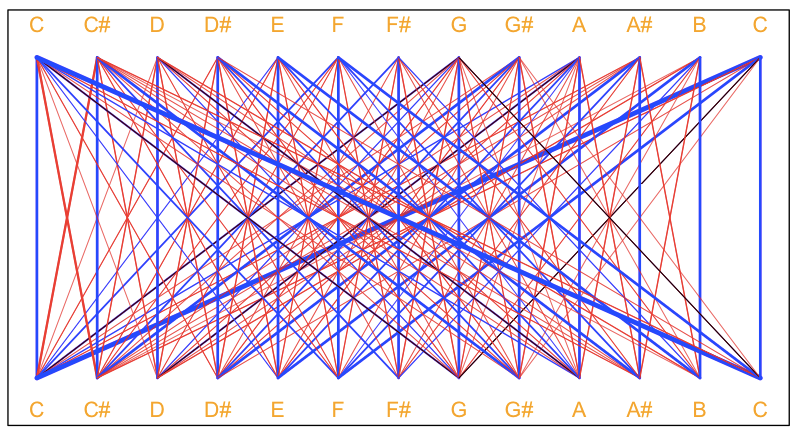} }
    \caption{Consonance/dissonance patterns obtained for the pythagorean (a),
    mean-tone (b), just major (c) and Werckmesiter scales (d).}   \label{fig:scales}
\end{figure*}

The consonance/dissonance results obtained for the \emph{pythagorean} scale are
shown in Figure~\ref{fig:scales}(a).  This scale is founded on the $2/3$ ratio interval,
corresponding to the pure perfect fifth, possibly the third most consonant interval
after the unison and octave.

Interestingly, the consonance/dissonance patterns turn out to be very similar
to those obtained for the equal temperament, but the consonances related
to the major forth and fifth now alternates for almost every node, sometimes
vanishing or being exchanged for other intervals.

Figure~\ref{fig:scales}(b) presents the consonance/dissonance patterns obtained
for the \emph{mean-tone} temperament.  The presence of a substantially increased
number of consonances can be readily observed.  For instance, most of the intervals
become consonant in the case of $C$.  The intervals involving $G$, an important tone, 
also tend to be consonant.  However, the aforementioned effect
is achieved at the expense of consonance uniformity.  Indeed, the intervals
based on $C\#$ are mostly dissonant, but they become slightly more consonant for
$D$.  It is also interesting to observe the variation of the intensities of the dissonances,
represented by the edges width.  Particularly strong dissonances are observed for
the intervals $D-D\#$ and $D\#-E$.  These result are in agreement with the fact
that the mean-tone will provide best results for the first modes, e.g. $C$-major
and $G$-major, becoming less suitable for further mode transpositions.

An even more consonant overall pattern is observed for the \emph{just major} scale,
depicted in Figure~\ref{fig:scales}(d).  Indeed, this results is similar to that observed
for the mean-tone case.

The consonance/dissonance patterns obtained for the intervals of the 
\emph{Werckmeister} temperament, shown in Figure~\ref{fig:scales}(e) seem
to be similar to those obtained for the equal temperament and pythagorean.

\section{Scales as Graphs}

\begin{figure*}
\centering{ 
        Equal  \\
        \includegraphics[width=6cm]{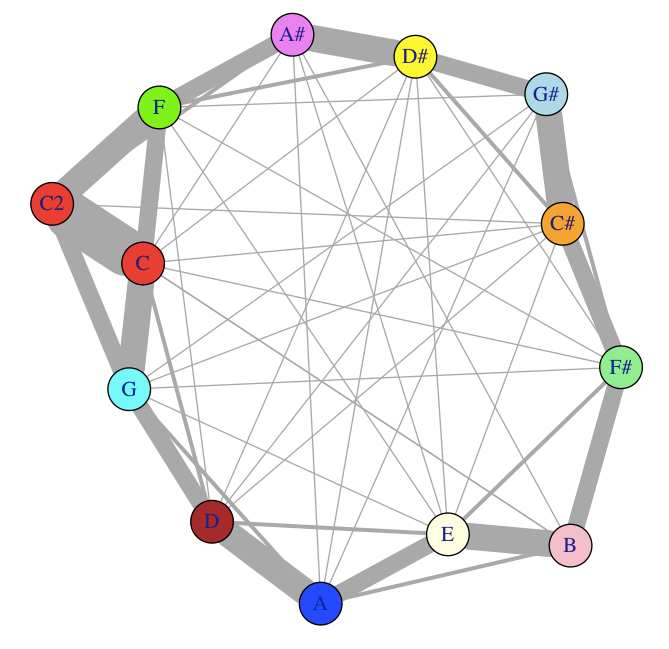} \\ Pythagorean:  \hspace{5cm} Meantone: \\
        \includegraphics[width=6cm]{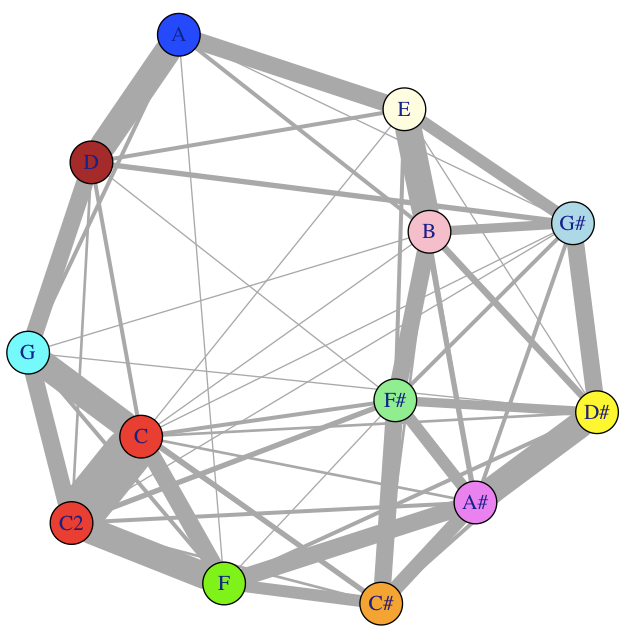}   \hspace{2cm}
        \includegraphics[width=6cm]{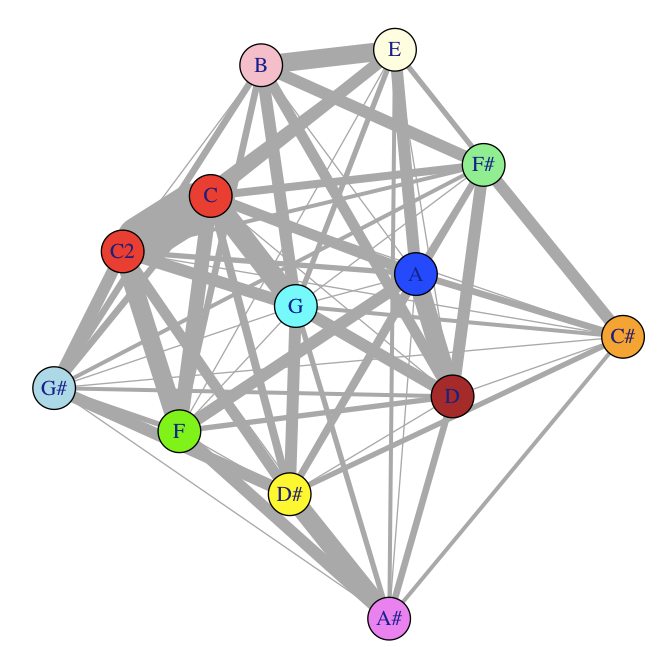}  \\  \vspace{0.5cm} 
             Just Major: \hspace{5cm}  Werckmeister: \\
        \includegraphics[width=6cm]{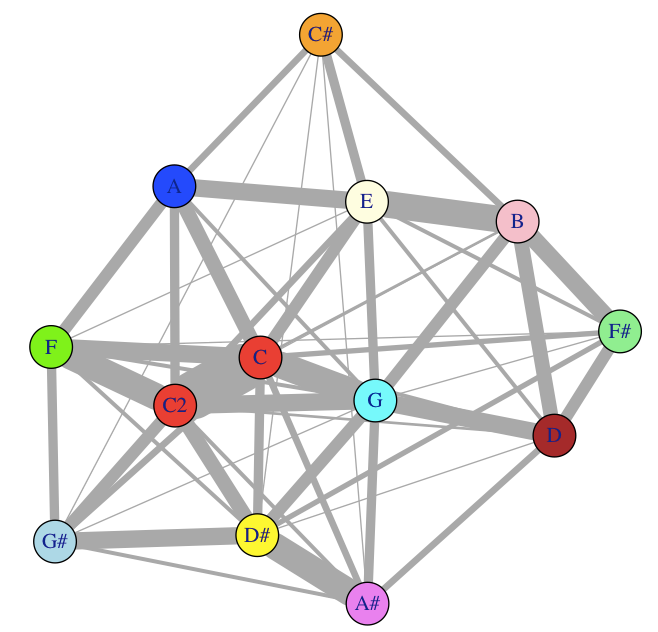}  \hspace{2cm}
        \includegraphics[width=6cm]{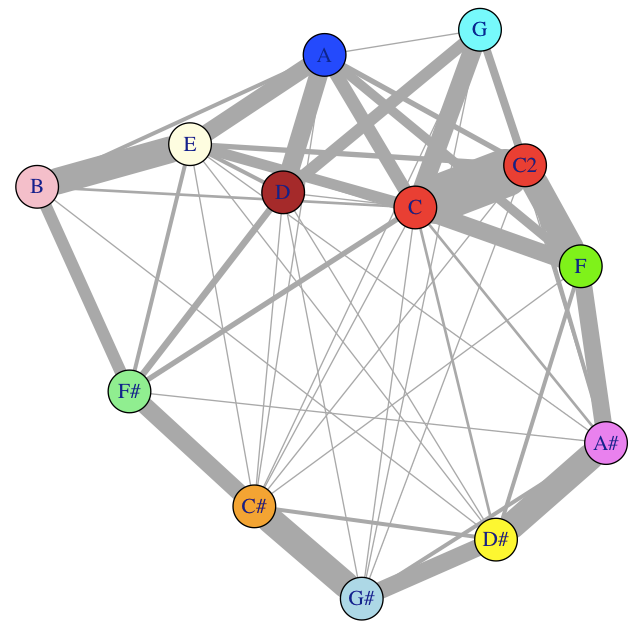}     
    \caption{The \emph{consonance patterns} obtained for the considered
      temperaments are shown in this figure, allowing further insights about
      the respective properties and relationships, as discussed in the main text.
      The intensities of the consonances are reflected in the edge widths.}  \label{fig:graphs}}
\end{figure*}

\begin{figure*}
\centering{ 
        Equal  \\
        \includegraphics[width=6cm]{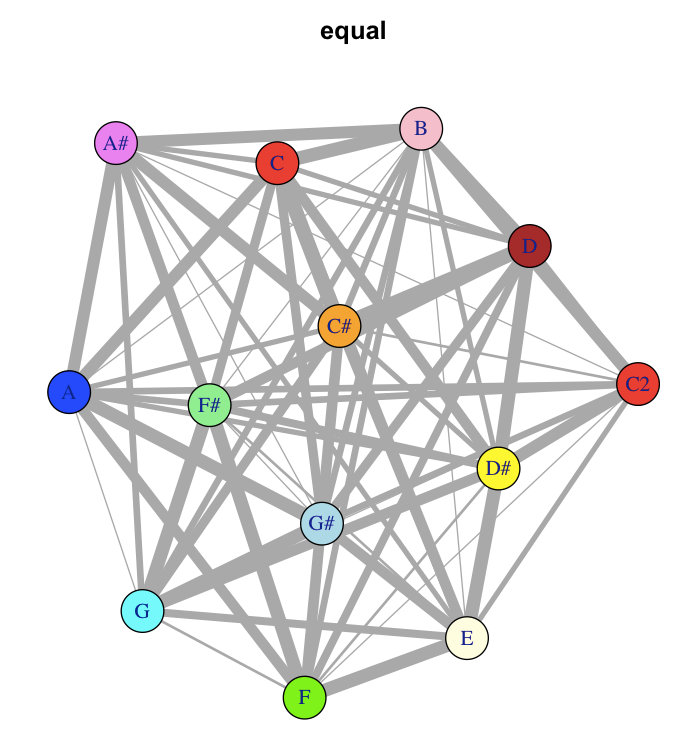} \\ Pythagorean:  \hspace{5cm} Meantone: \\
        \includegraphics[width=6cm]{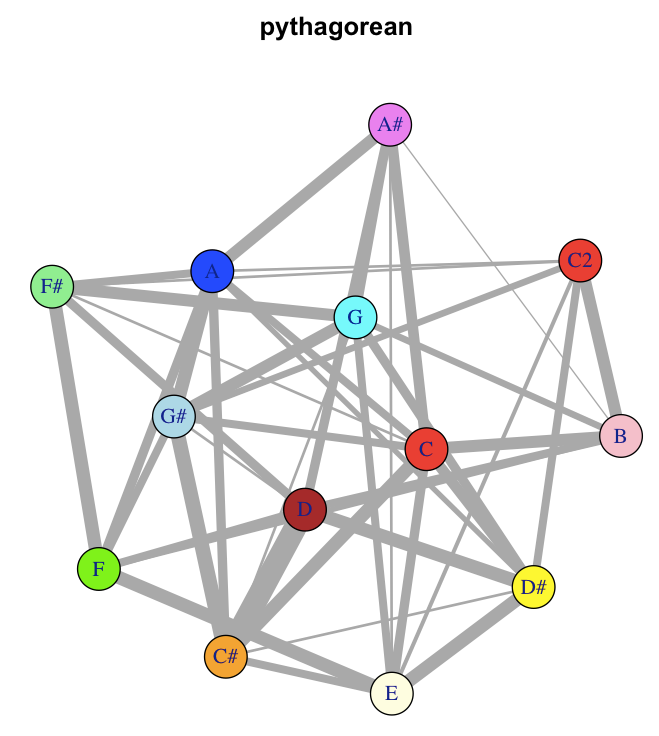}   \hspace{2cm}
        \includegraphics[width=6cm]{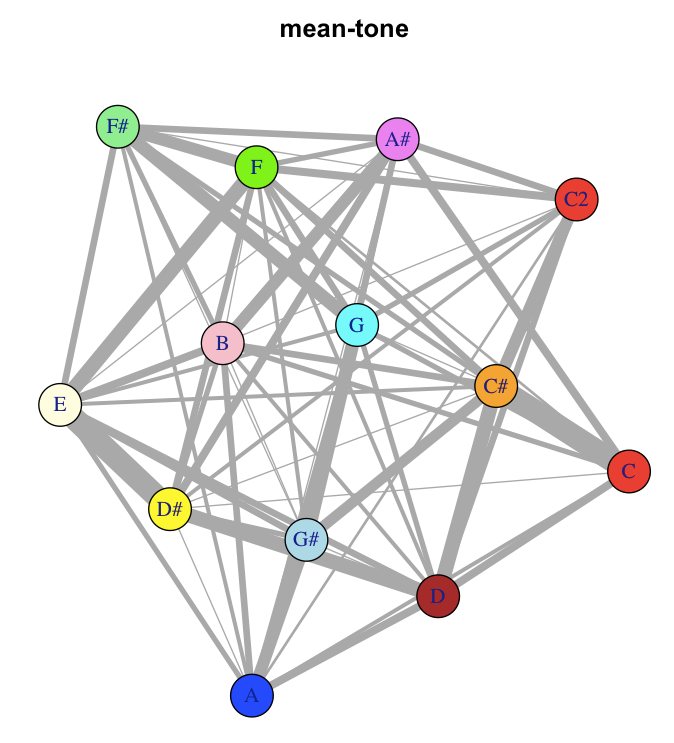}  \\  \vspace{0.5cm} 
             Just Major: \hspace{5cm}  Werckmeister: \\
        \includegraphics[width=6cm]{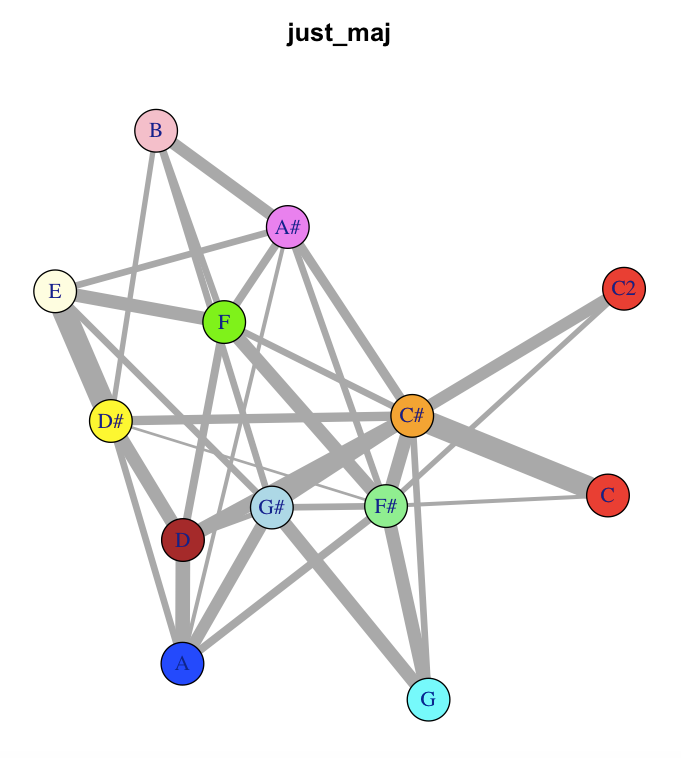}  \hspace{2cm}
        \includegraphics[width=6cm]{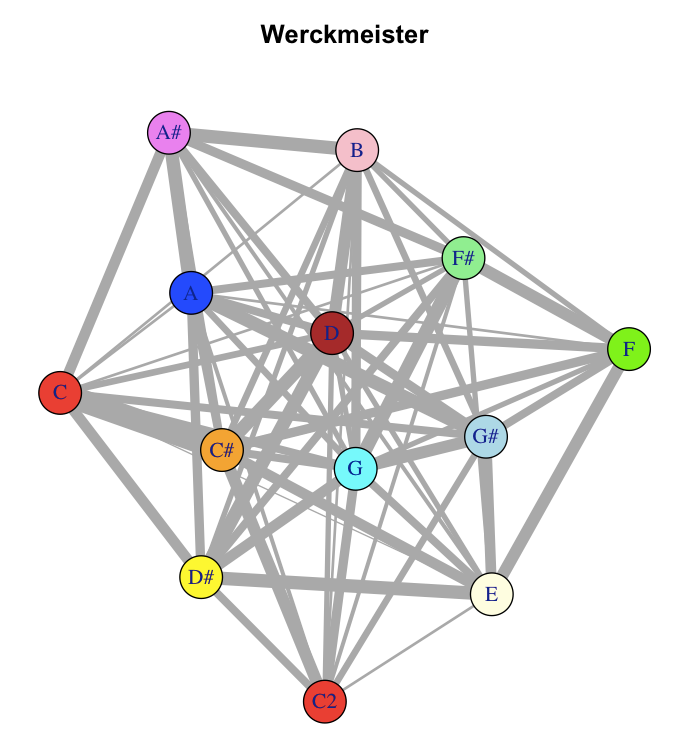}     
    \caption{The \emph{dissonance patterns} obtained for the considered
      temperaments are shown in this figure.  
      The intensities of the dissonances are reflected in the edge widths.
       Observe that the dissonances coexist with consonances, so that
       the same pair of edges can have meaningful consonance and
       dissonance connections.  However, these two features tend to
       be negatively correlated.}  \label{fig:graphs_diss}}
\end{figure*}

Visualizing the consonances as graphs (e.g.~\cite{CostaVisualiz}) where each
node corresponds to a single node can provide further insights
about the properties and relationships between diverse temperaments.  
Figures~\ref{fig:graphs} and~\ref{fig:graphs_diss} present, respectively, 
the consonance and dissonance graphs regarding the results discussed
in the previous section.

Remarkably, this type of visualization allows a nice complementation of the
information provided by the previously considered bipartite interrelationships.
Now, it becomes evident that we have two main groups of scales: one
similar to the equal temperament (also including the pythagorean and Werckmeister),
and the other corresponding to the mean-tone and just major.  

The fact that the mean-tone and just major scales tend to promote overall consonance
is directly reflected in the greater interconnectivity exhibited by the respective graphs.
The more uniform interval distributions underlying the other three types of scales define
a chain of consonances between several notes, with a cluster of connections being
observed between $C$, $G$, $C2$, and $F$.  

In the Werckmeister temperament,
this cluster also includes $A$ and $E$.  The consonances between non-adjacent
notes in these three temperaments tend to be little consonant, with the
pythagorean scale providing possibly improved consonance in this case.

\section{Triads and Harmony}

Harmony, the art of combining simultaneous sounds, is largely related to
the consonance/dissonance properties characteristic of the involved intervals.  
The minor (major) \emph{ds},
corresponding to the basis tone plus the minor (major) triad and fifth,
provides one of the most important foundations of traditional harmony.

\begin{figure}
\centering{
\includegraphics[width=8cm]{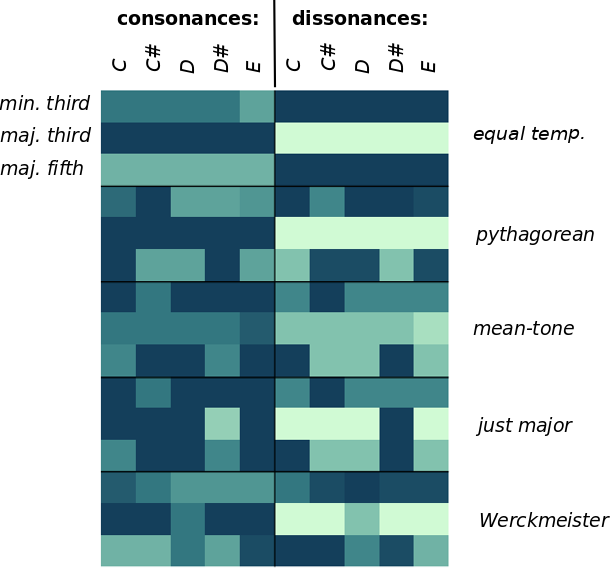}  
\caption{The consonance (left columns) and dissonances (right columns),
     with respective intensity levels coded from dark to bright,
     obtained for the minor and major triads considering the five temperaments.}
\label{fig:triads}}
\end{figure}

Let's also apply the developed simple model of consonance/dissonance to 
characterize the first minor and major triads obtained for the considered
five temperament types.  Figure~\ref{fig:triads} present the respective
consonance (left part of the table) and dissonance (right portion) 
levels.

The intrinsic uniformity of the equal temperament approach is evident 
from these results, characterized by constant horizontal intensities.
The other cases are characterized by the respective triads having varying
consonance/dissonance patterns.   This would be reflected in a different
effects while transposing or modulating along a music piece played according to these
temperaments.

\section{Electronic Amplification}

Let's us now illustrate the application of the suggested consonance/dissonance model to 
the characterization of electronic amplification 
(e.g.~\cite{boylestad:2008,jaeger:1997,lin:2017,costafeed:2017}).  Basically, an electronic amplified applies
a mapping $y(t) = f \left[ s(t) \right]$ of the input signal $s(t)$.  Ideally, in order to avoid distortion,
the mapping $f \left[ s(t) \right]$ should correspond to be linear 
(e.g. $f\left[ s(t) \right] = A s(t)$, for some gain $A$)

However, it is virtually impossible to obtain a completely linear amplification, so that we have
to cope with some level of nonlinearity.   Here, we consider quadratic nonlinearity, 
i.e.~$ f \left[ s(t) \right]  = \left[  s(t) \right] ^2$.  Unlike what happens in linear amplification,
new frequencies corresponding to sums and subtractions of the original frequencies, are
obtained as part of the resulting amplified signal $y(t)$.  In case the input signal contains
three frequencies $\omega_1$, $\omega_2$, and $\omega_3$, as shown in
Equation~\ref{eq:st}, the output will be as given in Equation~\ref{eq:s2t}.

\begin{equation}  \label{eq:st}
    s(t) = a_1 cos(\omega_1 t) + a_2 cos(\omega_2 t) + a_3 cos(\omega_3 t) + a
\end{equation}

\begin{eqnarray}  \label{eq:s2t}
   s^2(t) = a_3^2 cos(2 \omega_3 t) + a_2^2 cos (2 \omega_2 t) + 2 a_1^2 cos(2 \omega_1 t)  + \nonumber \\
   + 2 a a_3 cos(\omega_3 t) + 2 a a_2 cos(\omega_2 t) + 2 a a_1 cos(\omega_1 t)   +   \nonumber \\
   + 2 a_2 a_3 cos((\omega_2+\omega_3) t) + 2 a_2 a_3 cos((\omega_2-\omega_3)t)   + \nonumber \\
   + 2 a_1 a_3 cos((\omega_1+\omega_3)t)  +  2 a_1 a_3 cos(\omega_1 -\omega_3)t)  + \nonumber \\
   + 2 a_1 a_2 cos ((\omega_1 + \omega_2))  + 2 a_1 a_2 cos((\omega_1 - \omega_2)t)  +  \nonumber \\
   + 5 a^2 + a_1^2 + a_2^2
\end{eqnarray}

\begin{figure}
\centering{
\includegraphics[width=8cm]{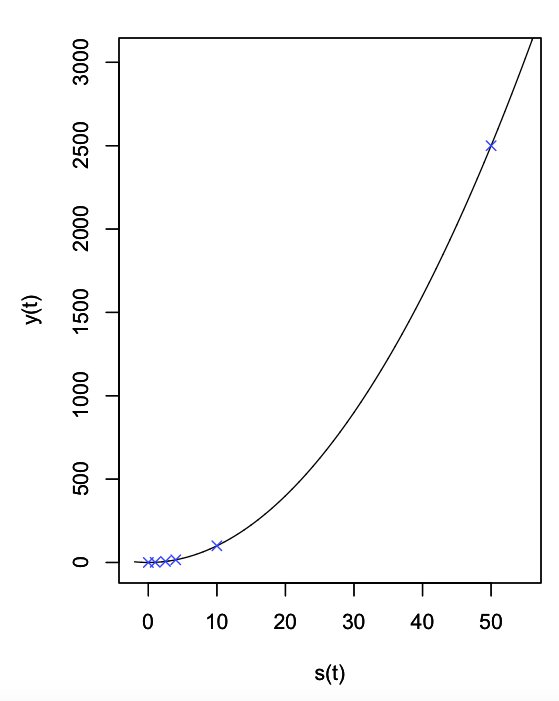}  
\caption{The quadratic nonlinear transfer function assumed in this section.  The blue marks identify
                the considered biasing $a = 0, 1, 2.5, 4, 10, 50$.  Observe that the linearity of the
                amplification increases with $a$.}
\label{fig:quadr}}
\end{figure}

The new frequencies induced by the nonlinearity provide new opportunities for obtaining consonance
and dissonance, so it is interesting to compare the effect of the nonlinearity on the resulting sound.
Figure~\ref{fig:quadr} depicts the quadratic amplification curve adopted in this section.  The level of
nonlinearity can be controlled by selecting different biasing intensities, so that the linearity
increases with this parameter, i.e. $L = a$.

\begin{figure*}
\centering{
\includegraphics[width=15cm]{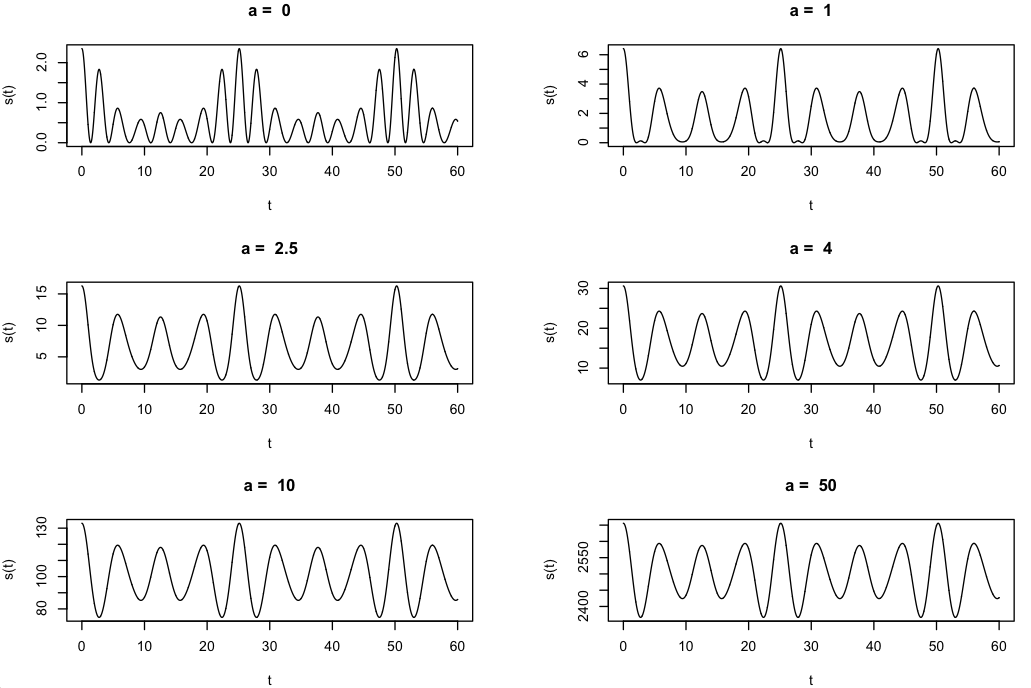}  
\caption{The sound signals resulting from the considered quadratic amplification for 
      increasing linearity levels defined by the biasing value $a$.  The result obtained for
      $a=50$ is almost identical to what would be obtained from a fully linear amplification.
      Though the effects of the quadratic nonlinearity are nearly imperceptible for
      $a = 4$ and $10$, the respective consonance graphs in Figure~\ref{fig:nonlingraphs}
      indicated substantial alterations of the respective consonances.}
\label{fig:nonlin}}
\end{figure*}

Figure~\ref{fig:nonlin} shows the output signals obtained for the input signal 
$s(t) = cos(\omega_1 t) + 1/3  \, cos(\omega_2 t) + 1/5 \, cos(\omega_3 t) + a$ considering the
three involved frequencies as corresponding to the just major fundamental, major third and
major fifth.

\begin{figure*}
\centering{ 
        \includegraphics[width=6cm]{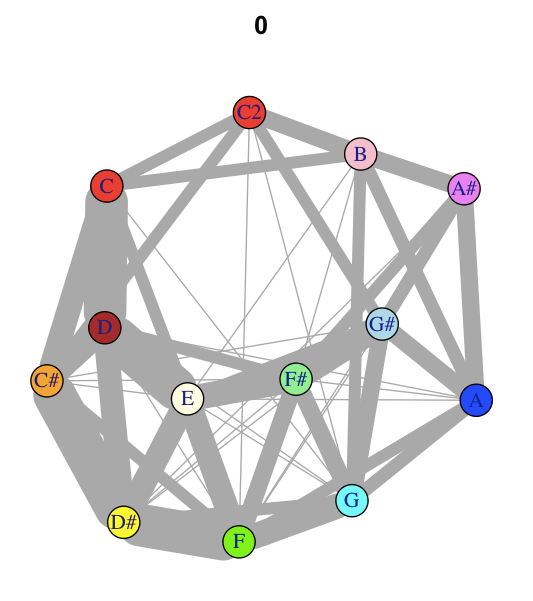} \hspace{2cm}
        \includegraphics[width=6cm]{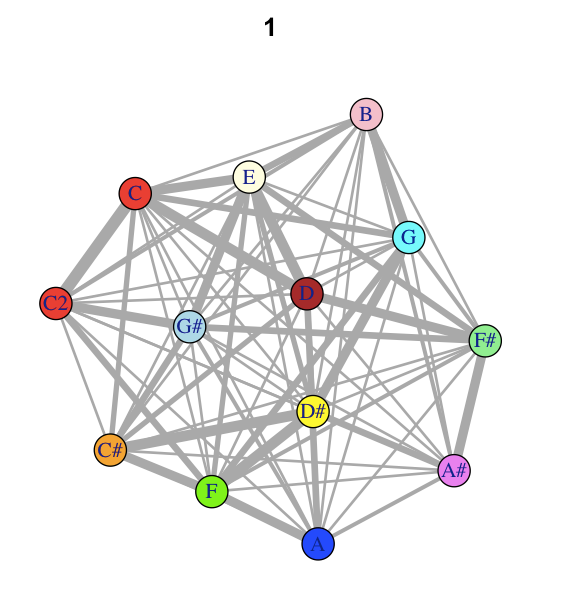}    \\  \vspace{0.5cm} 
        \includegraphics[width=6cm]{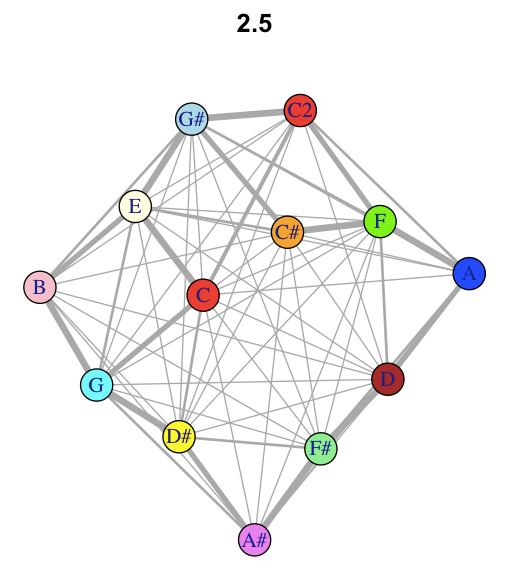}  \hspace{2cm}
        \includegraphics[width=6cm]{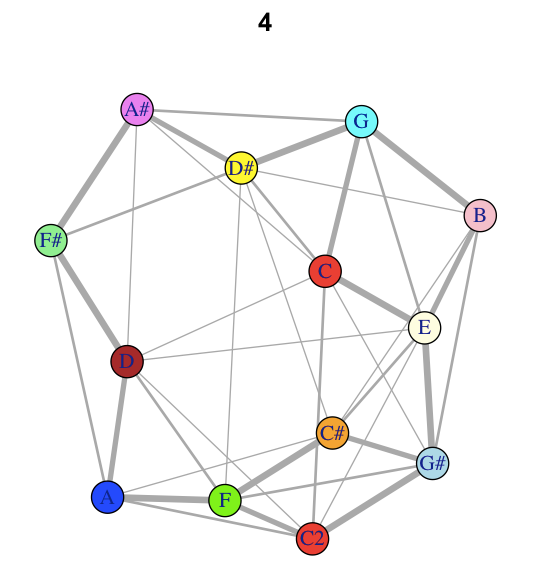}  \\ \vspace{0.5cm}
        \includegraphics[width=6cm]{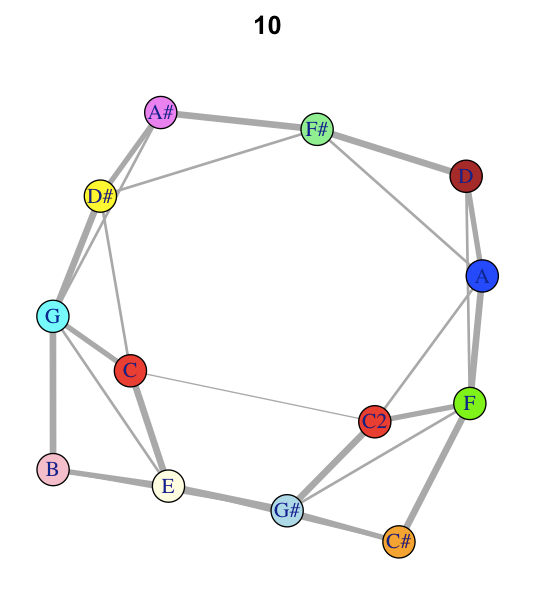}  \hspace{2cm}    
        \includegraphics[width=8cm]{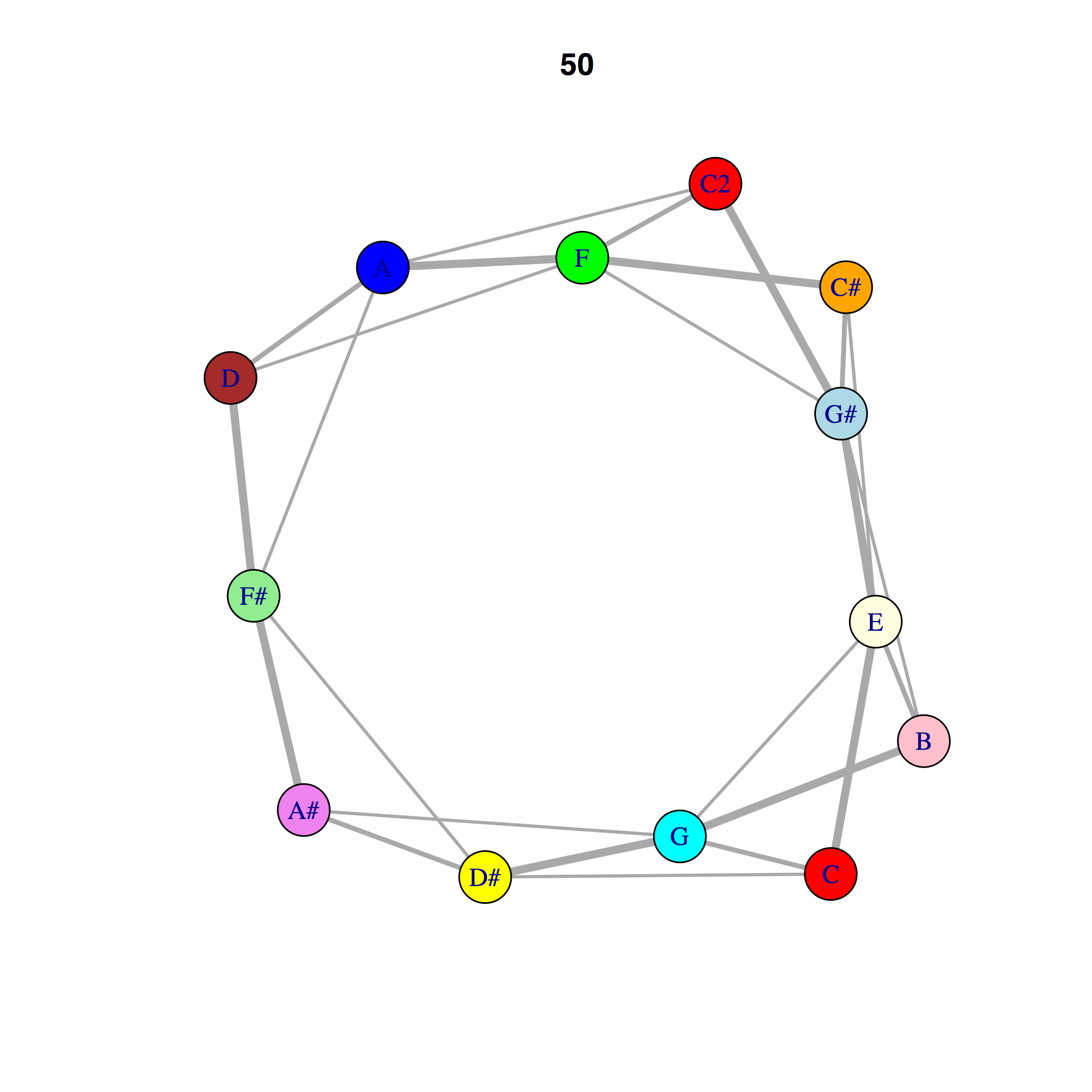}     
    \caption{The \emph{consonance patterns} obtained for several levels
    of nonlinear (quadratic) electronic amplification, ranging from highest
    ($L =0$ ) to lowest ($L=50$) considered nonlinearities.}  \label{fig:nonlingraphs}}
\end{figure*}

The effects of nonlinearity are evident in the first two or three cases (i.e. $L = 0, 1 and 2.5$),
and are much less perceptible in the other cases.   However, the consonance graphs shown
in Figure~\ref{fig:nonlingraphs} reveal that the effects of the nonlinearity proceed until the 
last case (i.e.~$L = 50$), where the resulting simulated sound is nearly identical to what
would be otherwise obtained by using a fully linear amplification.

If follows from these simulations that amplification nonlinearity can substantially
chance the consonance (similar results were observed for the dissonances, not
shown here) relationships between pairs of notes.  For instance, the interval 
$D-G$, which exhibit a considerable consonance originally, results disconnected
for the largest dissonance case $a=0$.  Even more pronounced effects are obtained
as a consequence of the introduction of several harmonics not present in the original
signal, which give rise to new consonances and dissonances.

\section{Concluding Remarks}

We have briefly addressed the interesting issues of consonance and dissonance
in sound and music, with special focus on Helmholtz's consonance approach.
A simple model was developed that allowed us to quantify the levels of consonance
and dissonance between two tones according to comparisons of their respective
partials.  

The potential of this approach was then illustrated with respect to
an analysis of temperaments and triads, and interesting results were obtained that
are often in agreement with what is commonly believed.  In particular, the 
visualization of consonance relationships estimated for five types of temperaments
resulted to be particularly useful for identifying specific features and comparing
different scales.  The potential of the suggested consonance/dissonance model
was also illustrated with respect to the characterization of the strong effects
that nonlinear electronic amplification can have on changing the patterns of
consonance and dissonance of the amplified sounds.

It should be kept in mind that the obtained results are preliminary and dependent
on the model parameter configurations.  Substantial more efforts need to be
developed in further assessing the simple proposed approach, and new results
should be reported in the future.

We believe that, to any extent,
this work motivated the potential of consonance/dissonance models
with respect to sound and music aspects.  

In addition to the study of temperaments
and triads, it would be also possible to consider a large number of possible
applications including:  provide subsidy to design of scales
and composition, musical instrument construction and characterization, 
acoustics, design and characterization of electronic amplifiers
considering other types of nonlinearities, and
filters, to name but a few possibilities.

It would be also promising to try to quantify the intrinsic sound consonance
of a single note by considering interval relationships between its harmonics
through an adaptation of the described modeling approach.

\vspace{0.7cm}
\textbf{Acknowledgments.}

Luciano da F. Costa
thanks CNPq (grant no.~307085/2018-0) for sponsorship. This work has benefited from
FAPESP grant 15/22308-2.  
\vspace{1cm}

\bibliography{mybib}

\begin{thebibliography}{10}

\bibitem{Kahn:2001}
C.~H. Kahn.
\newblock {\em Pythagoras and the Pythagoreans}.
\newblock Hackett Publishing, 2001.

\bibitem{indian_music}
N.~A. Jairazbhoy.
\newblock Harmonic implications of consonance and dissonance in ancient
  {I}ndian music.
\newblock {\em Pacific Review of Echnomusicology}, 2:28--51, 1985.

\bibitem{Laitz:2009}
S.~G. Laitz and C.~Bartlette.
\newblock {\em Graduate Review of Tonal Theory: A Recasting of Common-Practice
  Harmony, Form, and Counterpoint}.
\newblock Oxford University Press, 2009.

\bibitem{Stone:2018}
S.~C. Stone.
\newblock {\em Music Theory and Composition: A Practical Approach}.
\newblock Rowman and Littlefield Publ., 2018.

\bibitem{Christensen:1993}
T.~Christensen.
\newblock {\em Rameau and Musical Thought in the Enlightenment}.
\newblock Cambridge University Press, 1993.

\bibitem{Cunningham}
R.~E.~Cunningham Jr.
\newblock Helmholtz's theory of consonance.
\newblock
  \url{http://www.robertcunninghamsmusic.com/PDFs/Helmholtzs_Theory_of_Consonance.pdf}.
  Online; accessed 10-June-2019.

\bibitem{CostaCons}
L.~da~F.~Costa.
\newblock Modeling consonance and its relationships with temperament and
  harmony.
\newblock Researchgate, 2019.
\newblock
  \url{https://www.researchgate.net/publication/333675642_Modeling_Consonance_and_Its_Relationships_with_Temperament_and_Harmony_CDT-10}.
  Online; accessed 10-June-2019.

\bibitem{CostaModeling}
L.~da~F.~Costa.
\newblock Modeling: The human approach to science.
\newblock Researchgate, 2019.
\newblock
  \url{https://www.researchgate.net/publication/333389500_Modeling_The_Human_Approach_to_Science_CDT-8}.
  Online; accessed 03-June-2019.

\bibitem{CostaPhasor}
L.~da~F.~Costa.
\newblock Circuits, oscillations, and the {K}uramoto model as visualized by
  phasors.
\newblock Researchgate, 2019.
\newblock
  \url{https://www.researchgate.net/publication/333224636_Circuits_Oscillations_and_the_Kuramoto_Model_as_Visualized_by_Phasors_CDT-7}.
  Online; accessed 03-June-2019.

\bibitem{CostaVisualiz}
F.~N. Silva and L.~da~F.~Costa.
\newblock Visualizing complex networks.
\newblock Researchgate, 2018.
\newblock
  \url{https://www.researchgate.net/publication/328811693_Visualizing_Complex_Networks_CDT-5}.
  Online; accessed 03-June-2019.

\bibitem{boylestad:2008}
R.~L. Boylestad and L.~Nashelsky.
\newblock {\em Electronic Devices and Circuit Theory}.
\newblock Pearson, 2008.

\bibitem{jaeger:1997}
R.~C. Jaeger and T.~N. Blalock.
\newblock {\em Microelectronic Circuit Design}.
\newblock McGraw-Hill New York, 1997.

\bibitem{lin:2017}
F.N. Silva, C.~H. Comin, and L.~da~F. Costa.
\newblock Seeking maximum linearity of transfer functions.
\newblock {\em Rev. Sci. Instrums.}, 87(124701), 2016.

\bibitem{costafeed:2017}
L.~da~F. Costa, F.~N. Silva, and C.~H. Comin.
\newblock Negative feedback, linearity and parameter invariance in linear
  electronic.
\newblock {\em Electrical Engineering}, 99(3):1139, 2017.

\end{thebibliography}
\bibliographystyle{unsrt}
\end{document}